\documentclass[aps,prx,onecolumn,footinbib]{revtex4-2}
\usepackage{graphicx}
\usepackage{indentfirst}
\usepackage{physics}
\usepackage{braket}
\usepackage{float}
\usepackage{amsmath}
\usepackage{epstopdf}
\usepackage{footnote}
\usepackage{CJK}
\usepackage{esint}
\usepackage{color}
\usepackage[T1]{fontenc}
\usepackage{subfigure}
\usepackage{amsfonts}
\usepackage{footmisc}
\usepackage{scrextend}
\usepackage{multirow}
\usepackage[hyperfootnotes=false]{hyperref}
\usepackage[acronym]{glossaries}

\newacronym{QN}{QN}{quantum network}
\newacronym{QI}{QI}{Quantum information}

\usepackage[english]{babel}
\usepackage{url}
\usepackage{bm}
\usepackage{hyperref}
\definecolor{darkblue}{rgb}{0,0,0.5}
\hypersetup{
colorlinks=true,
linkcolor=black,
filecolor=blue,
citecolor=darkblue,  
urlcolor=black,
}

\newcommand{\etal}{{\it et al.}}
\newcommand{\PRL}{Phys. Rev. Lett.}

\def\be{\begin{equation}}
\def\ee{\end{equation}}
\def\ba{\begin{eqnarray}}
\def\ea{\end{eqnarray}}
\usepackage{bm}

\newcommand{\QZ}[1]{{{\textcolor{black}{#1}}}}

\begin{document}

\title{Quantum Internet under random breakdowns and intentional attacks}
 
\author{Bingzhi Zhang$^{1,2}$}
\author{Quntao Zhuang$^{1,3}$}
\email{zhuangquntao@email.arizona.edu}

\address{
$^1$Department of Electrical and Computer Engineering, University of Arizona, Tucson, Arizona 85721, USA
}
\address{
$^2$Department of Physics, University of Arizona, Tucson, AZ 85721, USA
}
\address{
$^3$James C. Wyant College of Optical Sciences, University of Arizona, Tucson, AZ 85721, USA
}

\begin{abstract}
Quantum networks will play a key role in distributed quantum information processing. As the network size increases, network-level errors like random breakdown and intentional attack are inevitable; therefore, it is important to understand the robustness of large-scale quantum networks, similar to what has been done for the classical counterpart---the Internet. For exponential networks such as Waxman networks, errors simply re-parameterize the network and lead to a linear decrease of the quantum capacity with the probability of error. 
The same linear decay happens for scale-free quantum networks under random breakdowns, despite the previously discovered robustness in terms of the connectivity. In presence of attack, however, the capacity of scale-free quantum networks shows a sharp exponential decay with the increasing attack fraction. Our results apply to quantum internet based on fibers for all kinds of quantum communications and provide implications for the future construction of quantum networks with regard to its robustness.
\end{abstract}
\maketitle
%
%
%
%
%

\section{Introduction}

The Internet enables networking between classical computers for information transmission and distributed information processing~\cite{lynch1996distributed,andrews2000foundations}. With the development of quantum information science, while various applications---quantum computing~\cite{Shor_1997}, quantum communication~\cite{gisin2007quantum} and quantum sensing~\cite{Giovannetti2004}---are being pursued, a \acrfull{QN}~\cite{kimble2008quantum,biamonte2019complex,wehner2018quantum,kozlowski2019towards,miguel2020genuine} is also emerging, with the ambitious goal to connect distant quantum devices via controlled quantum information transmission.

Quantum phenomena such as entanglement are usually fragile at a small scale. The same applies to communication links: a single quantum communication link is fragile to random \QZ{node or edge} breakdowns and attack. However, things can be different when there are more. For example, robust quantum phenomena persist in many-body systems despite the presence of local perturbations, giving rise to topological phases of matter~\cite{zeng2019quantum}. In this regard, it would be interesting to explore whether a \acrshort{QN}, with a large number of interconnected nodes and a complicated underlying dynamics induced by various protocols, can maintain an appreciable rate of end-to-end quantum communication in presence of random breakdown and attack.

In a classical network, the robustness of a network concerns its capability to remain fully-connected even when network errors occur. Pioneer studies~\cite{albert2000error,cohen2000resilience,cohen2001breakdown} have revealed the surprising robustness of the exponential networks (Erd\H{o}s-R\'enyi based models)~\cite{waxman1988routing,lakhina2003geographic} and scale-free networks~\cite{barabasi1999emergence,yook2002modeling} against node failures. However, scale-free networks become much more fragile than exponential networks when they undergo attack targeted at nodes with the highest degree of connection. 

In a \acrshort{QN}, different from a classical network, quantum information transmission rates are fundamentally limited by the links, regardless of the transmitter power. Indeed, the quantum capacity---the ultimate rate of quantum information transmission---for a fiber link decays exponentially with the distance~\cite{pirandola2009direct,takeoka2014fundamental,pirandola2017fundamental,pirandola2019end,azuma2016fundamental}. Therefore, even for a fully-connected \acrshort{QN}, the sustained rate can be very low. For this reason, quantitative characterizations beyond the connectivity is necessary to describe the communication capability of \acrshort{QN}s. Our earlier work~\cite{ourpaper} considers the typical end-to-end capacity~\cite{pirandola2019end} and study its growth with the density of nodes; We find a capacity threshold transition for exponential Waxman \acrshort{QN}s~\cite{waxman1988routing} and a saturation of capacity in scale-free \acrshort{QN}s~\cite{yook2002modeling}. To achieve an appreciable communication rate, the required density of nodes turns out to be orders-of-magnitude larger than the rough estimation based on connectivity analyses~\cite{brito2020statistical}.

In this work, we study the capability of \acrshort{QN}s to sustain an appreciable quantum capacity against various type of network errors---random node or edge breakdowns and attack targeted at nodes with the highest degree or total link capacity \QZ{\footnote{We do not study the attack on edges because that the measure of importance of an edge is not very clear and highly related to the node it is connected to.}}. To represent different possibilities of \acrshort{QN}s, we analyze both the Waxman and scale-free \acrshort{QN}s. 
The Waxman \acrshort{QN} has a capacity gradually decreasing with the increasing error under both random \QZ{(edge or node)} breakdown and attack, as errors merely continuously change network parameters while the network is still within the Waxman class. Surprisingly, the same continuous decay happens for scale-free \acrshort{QN}s under random \QZ{(edge or node)} breakdown, despite the previously discovered robustness in the connectivity~\cite{albert2000error,cohen2000resilience}. In presence of attack, however, the scale-free \acrshort{QN}s show an exponential capacity decay versus the attack fraction. This sharp decay quantitatively demonstrates the fragility of scale-free \acrshort{QN}s against attack and provides an alarm for the future designs of \acrshort{QN}s. \QZ{It also emphasizes the importance of taking robustness into consideration when designing quantum networks, as explored in Ref.~\cite{rabbie2020}.} Our results apply to all types of quantum communication through fibers, as the quantum capacity assisted by two-way communication, the secret key capacity and the entanglement distribution capacity are equal for bosonic loss channels~\cite{pirandola2019end}. 

\section{Network set-up}
As a large-scale \acrshort{QN} has not been built, we study the robustness of configurational models based on the existing classical fiber networks, which can be described by either the Waxman model~\cite{waxman1988routing} or the scale-free network~\cite{yook2002modeling}. The reason is that fiber networks are likely to be the bases of the infrastructure of a \acrshort{QN}, \QZ{as light is currently the main flying quantum information carrier (other platforms, such as solid-state~\cite{yamamoto2012} systems, are still not comparable)}. Without loss of generality, we choose $N_0$ randomly located nodes in a square of width $2R$ to construct the network. As \acrshort{QN}s require more sophisticated technology, errors can vary in types, but can all be characterized by a single parameter $p$. As for random breakdown, a node or a single edge is removed randomly with a probability $p$; while for intentional attack by degree/capacity, a fraction $p$ of nodes with the largest node degree/capacity are removed from the network. When a node is removed, all edges connecting to the node are also removed. Here we define the capacity of a single node as the total capacity of all of its edges, which takes the properties of the edges into account beyond the node degree.

To model optical communication, when two nodes located at $\bm x$ and $\bm x^\prime$ are connected, information transmission between them is enabled by a link modelled as a bosonic pure loss channel with a transmissivity \QZ{$\eta(\bm x, \bm x^\prime)=10^{-\gamma d(\bm x,\bm x^\prime)}$} at a state-of-the-art loss \QZ{$\gamma=0.02 \ {\rm km}^{-1}$}, where \QZ{$d(\bm x,\bm x^\prime)$} is the Euclidean distance.
In the Waxman model~\cite{waxman1988routing,lakhina2003geographic}, each pair of nodes is randomly connected with a probability decaying exponentially with the distance,
\QZ{
\begin{equation}
    \Pi_{\mathrm{W}}\left(\bm x, \bm x^\prime \right)= \beta_0 e^{-d(\bm x, \bm x^\prime)/\alpha L},
\label{prob_connect_waxman}
\end{equation}
}
where $L=2\sqrt{2}R$ is the maximum possible distance; The parameter $\beta_0$\footnote{In the rest of the paper, we will use subscript `$0$' to denote quantities in absence of errors. } determines the edge density and the constant $\alpha$ is chosen so that $\alpha L=226$km matches the US fiber networks~\cite{lakhina2003geographic}.
In the scale-free model~\cite{yook2002modeling}, the network is built up dynamically: when each node $\bm x$ is being added, it is connected to $m_0$ nodes out of all the previous added nodes. The probability of nodes $\bm x^\prime$ and $\bm x$ being connected obeys a power-law
\QZ{
\begin{equation}
\Pi_{\rm SF}\left(\bm x, \bm x^\prime \right)\propto D\left(\bm x\right)/d\left(\bm x, \bm x^\prime\right),
\label{prob_connect_scalefree}
\end{equation}
}
where \QZ{$D\left(\bm x\right)$} is the degree of the existing node $\bm x$, in contrast to the Waxman model's exponential decay.

\QZ{To transmit quantum information, quantum states with encoding are transmitted across each link. Despite various potential encoding schemes, the ultimate transmission rate (in qubits per channel use) of quantum information along an edge $E_{\bm{x},\bm{x}^\prime}$ (edge capacity)~\cite{pirandola2017fundamental} is 
\begin{equation}
    \mathcal{C}_E\left(E_{\bm x,\bm x^\prime}\right)=-\log_2\left(1-\eta(\bm x, \bm x^\prime)\right) = -\log_2(1-10^{-\gamma d(\bm x, \bm x^\prime)}),
\end{equation}
where $\eta$ is the transmissivity mentioned above. We also introduce the capacity of a node as the sum of the capacity of edges that connect to node $\bm x$ as
$\mathcal{C}_N(\bm x) = \sum_{\bm x^\prime \in \mathcal{N}(\bm x)}\mathcal{C}_E\left(E_{\bm x,\bm x^\prime}\right)$ where $\mathcal{N}(\bm x)$ are neighbors of node $\bm x$. Consider a classical network corresponding to a \acrshort{QN}, with the same set of nodes and edges, where the weight of an edge equals its edge capacity. In this regard, the problem to solve the end-to-end capacity in a \acrshort{QN} is equivalent to the minimum cut problem of the corresponding classical weighted graph~\cite{pirandola2019end}. We define a cut $\mathbb{U}_{\bm x, \bm x^\prime}$ as the set of all edges that will disconnect the two nodes $\bm x, \bm x^\prime$ after removal. Similar to the node capacity, we can define the cut capacity as the sum over edge capacities of edges in the cut, i.e., $C_U(\mathbb{U}_{\bm x, \bm x^\prime}) \equiv  \sum_{E_{\bm x, \bm x^\prime} \in \mathbb{U}_{\bm x,\bm x^\prime}} \mathcal{C}_E\left(E_{\bm x,\bm x^\prime}\right)$. Then the end-to-end capacity between nodes $\bm x, \bm x^\prime$ is given by the minimum cut capacity~\cite{pirandola2019end}
\begin{equation}
    \mathcal{C}(\bm x, \bm x^\prime) = \min_{\mathbb{U}_{\bm x,\bm x^\prime}} C_U(\mathbb{U}_{\bm x, \bm x^\prime}),
\end{equation}
where one optimizes over all cuts $\{\mathbb{U}_{\bm x, \bm x^\prime}\}$.
}



\QZ{
In the previous work~\cite{ourpaper}, we study the transition of average end-to-end capacity of two typical \acrshort{QN}s, Waxman and scale-free \acrshort{QN}s with respect to node density. For Waxman \acrshort{QN}s, we find that the average end-to-end capacity increases linearly with its node density, $\braket{\mathcal{C}}\propto \rho$, and thus there exists a critical density $\rho_c$ such that the average capacity $\braket{\mathcal{C}} = 1$; while for scale-free \acrshort{QN}s $\braket{\mathcal{C}}$ increases with the density of nodes but quickly reaches a constant for a fixed size $R$.
}

\QZ{In this paper, we consider the robustness of reliable end-to-end quantum communication against perturbations include breakdown and attack. Considering the different transitions for Waxman and scale-free \acrshort{QN}s, we choose the initial operating point of the two types of \acrshort{QN}s (initial node density (Waxman) and nodes number (scale-free)) prior to the perturbations accordingly to guarantee an appreciable end-to-end capacity $\braket{\mathcal{C}_0}\equiv \braket{\mathcal{C}(\bm x,\bm x^\prime)}$ when averaged over random node pairs $\bm x,\bm x^\prime$}~\cite{ourpaper}. For the Waxman \acrshort{QN}, we choose a density of nodes \QZ{$\rho_0\equiv N_0/4R^2 \simeq 4\times 10^{-4} {\rm km}^{-2}$} and $\beta_0=1$ such that $\braket{\mathcal{C}_0}\simeq 1$~\cite{ourpaper}; For the scale-free \acrshort{QN}, \QZ{$N_0=\lfloor10^{3.6}\rfloor=3981$} is chosen to be large enough so that $\braket{\mathcal{C}_0}$ saturates to a constant. With an average degree $2m_0=6$, the corresponding saturated $\braket{\mathcal{C}_0}\simeq 3, 0.4, 0.03$ for size $R = 40, 160, 400$ km~\cite{ourpaper}. \QZ{As the Waxman and scale-free \acrshort{QN}s behave in different ways with the increase of the density of nodes, our goal here is to understand their robustness at their own operating points.}

\begin{figure}
    \centering
    \includegraphics[width = 0.8\textwidth]{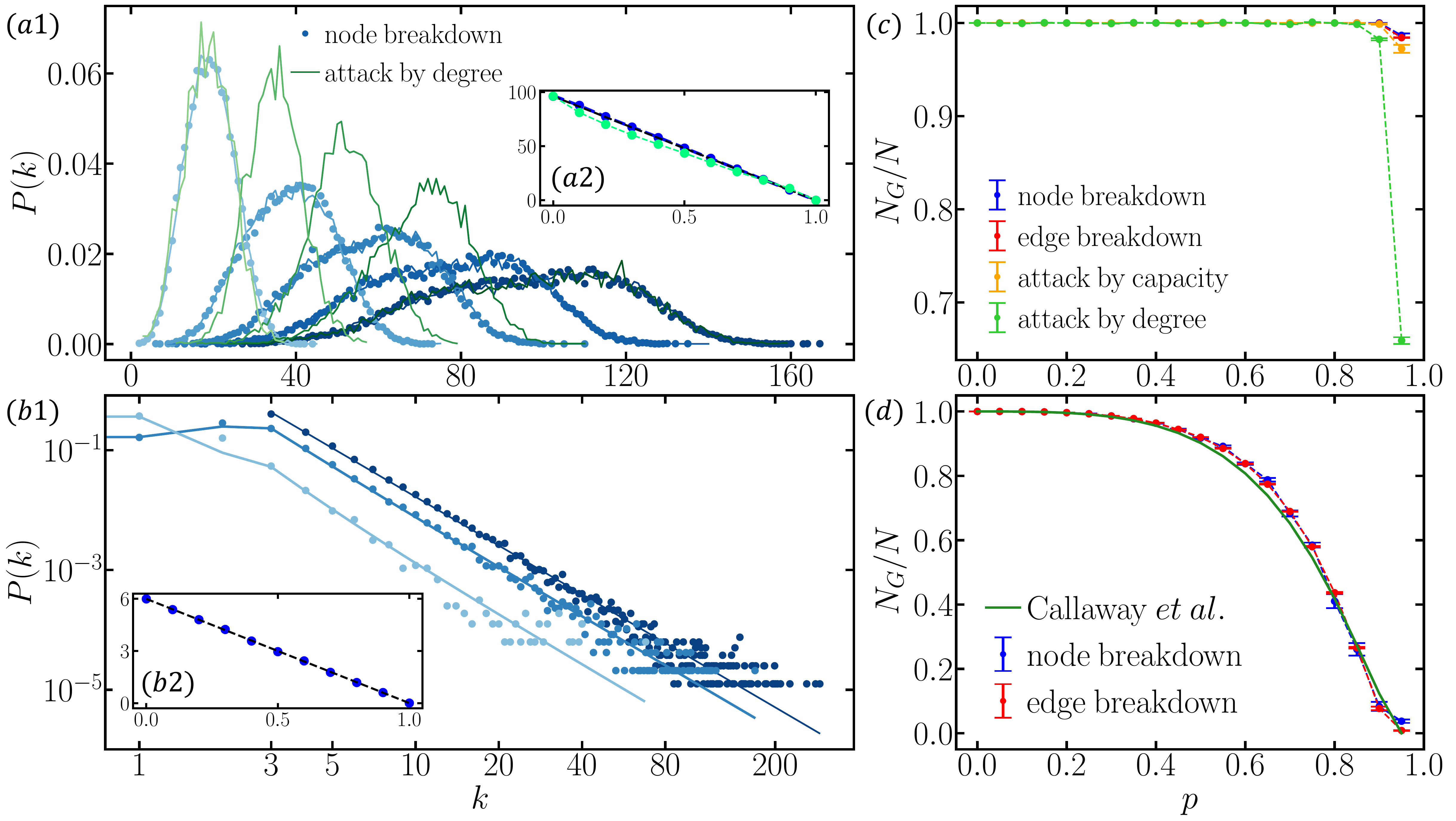}
    \caption{
    (a1) Degree distribution of Waxman model ($\alpha=0.1$) under random node breakdown (blue dots) and Waxman model with $N=N_0(1-p)$ nodes (blue curves). 
    Green curves represent $P(k)$ for Waxman model under attack by degree. 
    Dots and curves from dark to light represent $p=0, 0.2, 0.4, 0.6, 0.8$. 
    (b1) Degree distribution of scale-free networks under random node breakdown (dots) and analytical results (curves) with $R = 40$. Dots and curves from dark to light represent $p=0, 0.4, 0.8$. 
    Insets (a2), (b2) show the average degree $\braket{k}$ correspondingly and black dashed lines are $\braket{k_0}(1-p)$.  
    Relative size of the giant component in the Waxman model (c) and scale-free network (d). 
    Blue dots for random node breakdown and red for edge breakdown. 
    Orange and green dots in (c) for attack by capacity and degree separately. 
    The green curve in (d) is the result from \cite{callaway2000network}.}
    \label{fig:degree}
\end{figure}

\section{Waxman \acrshort{QN}s}
As the Waxman model is homogeneous in the thermodynamic limit---the degree distribution is centered at its mean without a long tail, we expect the Waxman model under random node breakdown to be equivalently reduced to a network with $N = N_0(1-p)$ nodes randomly selected from the initial $N_0$ nodes. We confirm the intuition by the degree distribution $P(k)$ in Fig.~\ref{fig:degree}(a) for the case of random node breakdown. The same applies to random edge breakdown and attack by capacity, as we explain in the following.
Under random edge breakdown, while the number of nodes $N_0$ is unchanged, the number of edges $|\mathcal{E}_0|$ of the initial Waxman model decreases to $|\mathcal{E}| = |\mathcal{E}_0|(1-p)$; As the parameter $\beta_0$ in Eq.~\ref{prob_connect_waxman} determines the edge density of the network \cite{waxman1988routing}, the Waxman model under random edge breakdown is equivalent to a Waxman model with edge connection probability
\QZ{
\begin{equation}
\Pi_{\rm W}^{\rm E}(\bm{x},\bm{x}^\prime) = \beta_0(1-p)e^{-d(\bm{x},\bm{x}^\prime)/\alpha L},
\label{eq:edge_probability}
\end{equation}
}
leading to an identical degree distribution to the case of node breakdown \QZ{(see Appendix A.)}~\cite{cohen2000resilience, martin2006random}. Here the superscript `E' denotes edge removal.
\QZ{Since the Waxman model is dense and its degree distribution shows a peak, we expect the node capacity to be roughly uniform and thus the attack by capacity is equivalent to random node breakdown, as confirmed numerically in Appendix A.}

The critical probability for the disappearance of giant components in the Waxman model is close to unity (Fig.~\ref{fig:degree}(c)), so there always exists a giant component in the Waxman model in the parameter region of interest. 

\begin{figure}
    \centering
    \includegraphics[width = 0.6\textwidth]{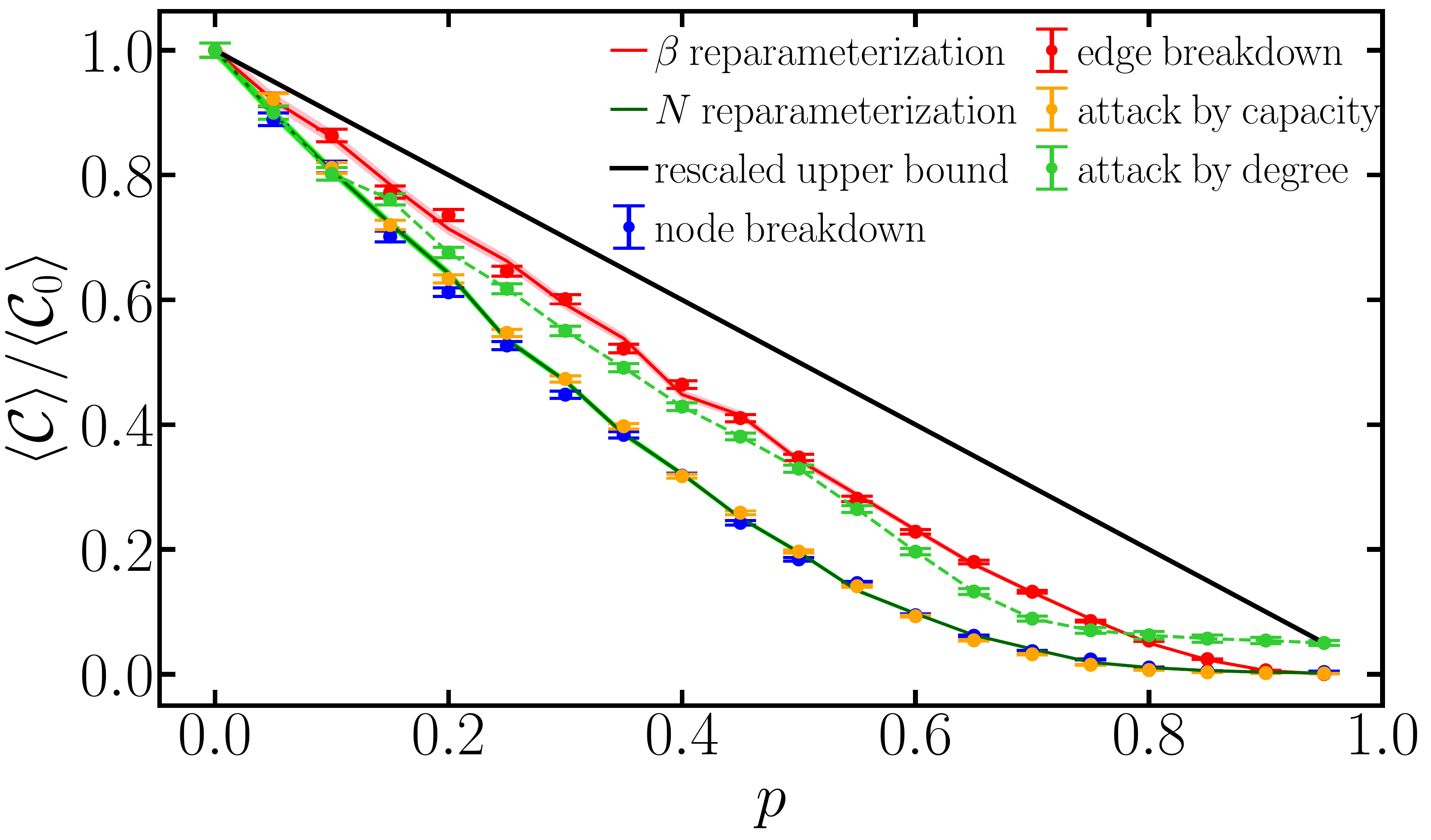}
    \caption{Normalized average end-to-end capacity $\braket{\mathcal{C}}/\braket{\mathcal{C}_0}$ of Waxman \acrshort{QN}s with $\alpha = 0.1$ under random node breakdown (blue), edge breakdown (red), attack by node capacity (orange) and attack by node degree (green) vs the error parameter $p$. \QZ{Dark green and red solid curves represent $\braket{\mathcal{C}}/\braket{\mathcal{C}_0}$ for Waxman \acrshort{QN}s with $N = N_0(1-p)$ nodes and $\beta = \beta_0(1-p)$, equivalent ones for node breakdown/attack by capacity and edge breakdown separately. Light color areas correspond to the error bars.} The black line represents the rescaled upper bound $\braket{\mathcal{C}}/\braket{\mathcal{C}_0} = 1-p$ for random breakdown and attack by capacity. $\braket{\mathcal{C}}/\braket{\mathcal{C}_0}$ of Waxman \acrshort{QN}s with other $\alpha$ under all cases except attack by degree also agrees with the one with $\alpha = 0.1$.
    }
    \label{fig:waxman_C}
\end{figure}

Now we proceed to analyze the capacity transition of the Waxman \acrshort{QN}s under breakdown or attack.
As shown in Fig.~\ref{fig:waxman_C}, the rescaled ensemble-averaged end-to-end capacity $\braket{\mathcal{C}}/\braket{\mathcal{C}_0}$ transits from unity to close to zero gradually for all cases, with a linear trend for $p$ being not too large. 
The cases of random node breakdown (blue dots) and attack on nodes by capacity (orange dots) overlaps, as predicted by the degree distribution.
However, despite the degree distribution also being the same for the case of edge breakdown, the quantum capacity (red dots) decays slower compared to the other cases. \QZ{This is because during edge breakdown, the edge density parameter $\beta = \beta_0(1-p)$ is reduced and leads to a slower linear decay of $\braket{\mathcal{C}}$ (see Appendix C.1), where we can see that the transition of $\braket{\mathcal{C}}$ vs $\beta$ is different from $\braket{\mathcal{C}}$ vs $\rho$ while $\beta = \beta_0 = 1$ is kept constant in the other cases. It also indicates that the agreement in degree distribution is only necessary but not sufficient condition for identical capacity.} As predicted from the degree distribution, the statistical equivalent Waxman \acrshort{QN} provides an accurate description of the transition of capacity in most cases, as the solid lines (orange, blue, red) produced from re-parameterized Waxman \acrshort{QN}s agree well with the numerical results (orange, blue, red dots) \QZ{(see Appendix C.1)}. 

As the degree distribution under attack by degree disagrees with the others, \QZ{we expect} the transition of the capacity under attack by node degree (green dots) to deviate from the other cases, as shown in Fig.~\ref{fig:waxman_C}. 
Note that when $p$ is close to unity, $\braket{\mathcal{C}}/\braket{\mathcal{C}_0}$ decays to a constant, due to small connected components remaining connected (see Appendix).

To understand the gradual transition, we resort to an upper bound of the node capacity $\mathcal{C}_N(\bm x)$---the total capacity of edges connecting to a node~\cite{ourpaper}
\begin{equation}
\label{eq:waxman_C_bound}
\braket{\mathcal{C}}\le \braket{\mathcal{C}_N(\bm x)} = \left(1-p\right) \zeta_{\rm W} \rho_0,
\end{equation}
where the constant $\zeta_{\rm W}$ is given by properties of a bosonic pure loss channel (see Appendix)~\cite{ourpaper} and $1-p$ is a factor coming from re-parameterization (see Appendix). We emphasize that this upper bound holds for scenarios under node or edge breakdowns, and attack. As the upper bounds are not tight even for the $p=0$ case, we rescale the upper bound to enable the direct comparison of the trend with $p$ in Fig.~\ref{fig:waxman_C}. The rescaled upper bound (black) is indeed larger than the rest; 

\section{Scale-free \acrshort{QN}s}
As scale-free \acrshort{QN}s show drastically different behaviors under random breakdown and under attack, we address breakdown and attack separately. In both cases, capacity decay will follow after the characterizations of classical properties.

\subsection{Random breakdown}
\label{sec:rb}
The degree distribution $P(k)$ of scale-free networks under random node or edge breakdown turns out to be identical and deviates from the initial power-law~\cite{stumpf2005subnets} (see Fig.~\ref{fig:degree}(b)), as analytically solved in Refs.~\cite{cohen2000resilience, martin2006random}. The average degree $\braket{k} = 2m_0(1-p)$ decreases linearly, same as the Waxman model. The change of the size $N_G$ of the giant component relative to the size $N$ of the broken network shrinks continuously with $p$ in the same way for both random node and edge breakdowns~\cite{callaway2000network}, and approaches zero when $p\sim 1$~\cite{cohen2000resilience}, as shown in Fig.~\ref{fig:degree}(d).


With the classical properties well-understood, now we derive analytical results for the quantum capacity transition.
For a scale-free \acrshort{QN} with $N_0$ nodes under random node or edge breakdown with probability $p$, typically $N_G$ nodes are within a giant component, and the rest are isolated in various tiny clusters. Because the capacity between nodes from different clusters is zero and the contribution from tiny clusters is small, the leading order contribution to the average end-to-end capacity $\braket{\mathcal{C}}$ comes from the cases when two nodes are chosen from the giant component.
\QZ{As $p$ increases, the \acrshort{QN} becomes fragmented, and the probability that a node does not belong to giant component $1-N_G/N$ becomes non-negligible, as shown in Fig.~\ref{fig:degree}. Denote the size of each small cluster as $N_k$, each cluster contributes when both two nodes are inside the same cluster, leading to a contribution $\propto (N_k/N)^2$. The total contribution is $\propto\sum_k (N_k/N)^2$, which is small as the number of small clusters $K$ is large. This can be easily seen by assuming all $K$ small clusters having the same number of nodes, then $\sum_k [(1-N_G/N)/K]^2\sim (1-N_G/N)^2/K\ll1$ when $K$ is large. }
Therefore, we have the relationship 
\begin{equation}
\label{eq:yook_nodebreak_C_sol}
\braket{\mathcal{C}}\times \binom{N}{2}=\braket{\mathcal{C}_G}\times \binom{N_G}{2},
\end{equation}
between $\braket{\mathcal{C}}$ and the average end-to-end capacity $\braket{\mathcal{C}_G}$ of the giant component in the broken network. Here $N=N_0$ for edge breakdown and $N=N_0(1-p)$ for node breakdown.  Both $\braket{\mathcal{C}}$ and $\braket{\mathcal{C}_G}$ are upper-bounded by $ 2(1-p)m_0\zeta_{\rm SF}$, where $\zeta_{\rm SF}$ is a constant determined by the rate-loss trade-off~\cite{ourpaper} (see Appendix) and the factor $2(1-p)m_0$ is the approximate average degree. However, combining Eq.~\ref{eq:yook_nodebreak_C_sol} and the upper bound for $\braket{\mathcal{C}_G}$ gives a better upper bound of $\braket{\mathcal{C}}$ as
\begin{equation}
\label{eq:yook_nodebreak_C_bound}
\braket{\mathcal{C}} \leq 2m_0(1-p)\zeta_{SF} \frac{\binom{N_G}{2}}{\binom{N}{2}} \propto (1-p)\left(\frac{N_G}{N}\right)^2.
\end{equation}
We see that the size of the giant component in Fig.~\ref{fig:degree}(d) comes in, directly connecting the quantum capacity to the giant component transition.
\begin{figure}[t]
    \centering
    \includegraphics[width = 0.7\textwidth]{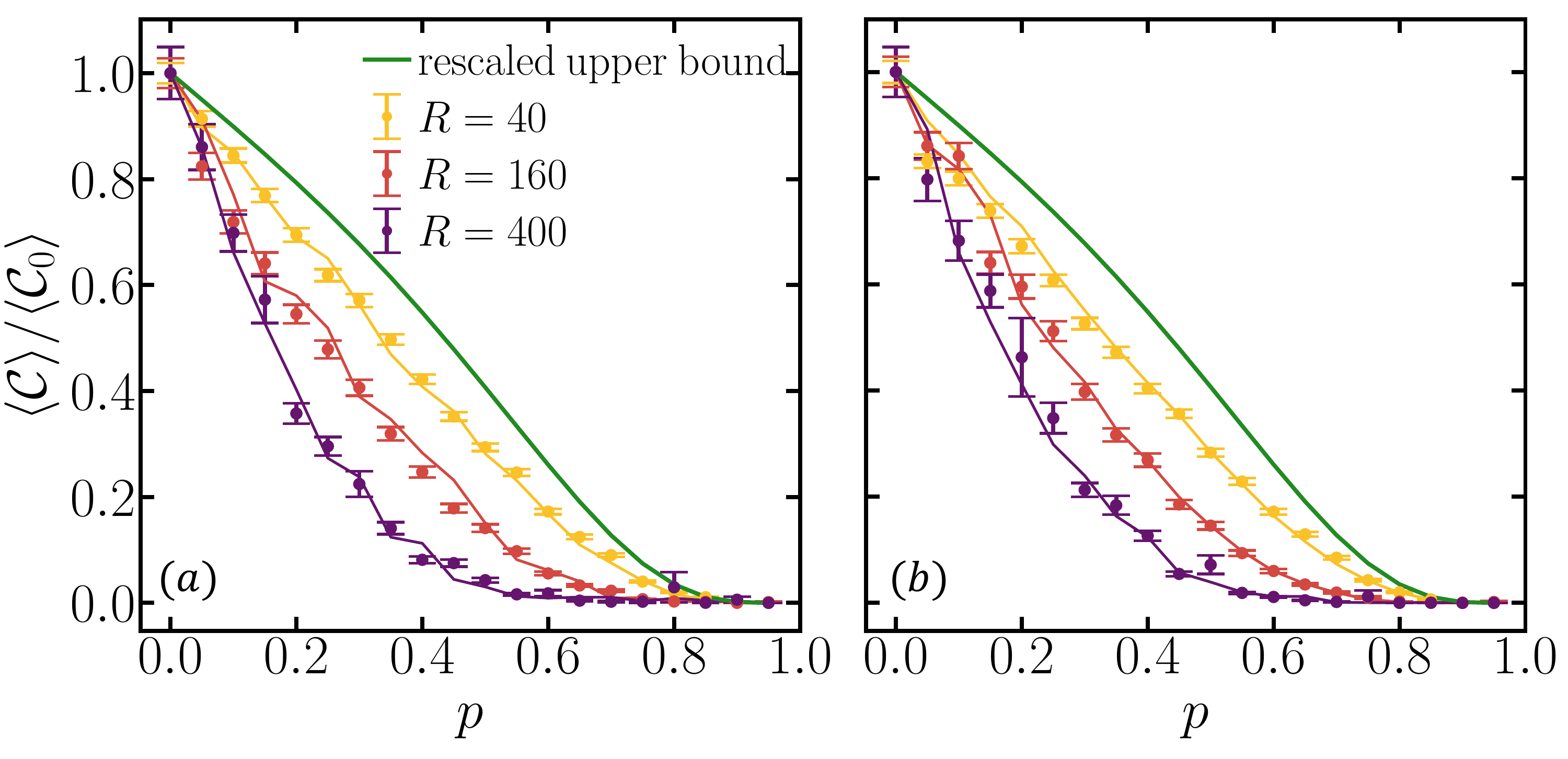}
    \caption{Normalized average end-to-end capacity $\braket{\mathcal{C}}/\braket{\mathcal{C}_0}$ vs $p$ of scale-free \acrshort{QN}s under random node (a) and edge (b) breakdown with $R = 40, 160, 400$. Subfigure (b) shares the same legend as in (a). Solid curves with the same color as dots represent the analytical solution in Eq.~\eqref{eq:yook_nodebreak_C_sol} in (a), (b) correspondingly. Green curves in (a), (b) represent the rescaled upper bound in Eq.~\eqref{eq:yook_nodebreak_C_bound} separately.
    }
    \label{fig:yookBreak_C}
\end{figure}
As the giant component ratio $N_G/N$ are identical for both node and edge breakdowns (see Fig.~\ref{fig:degree}(d)), the upper bound turns out to be identical under both node or edge breakdown for scale-free \acrshort{QN}s. We plot the numerical simulation results as well as the exact solution of the normalized average end-to-end capacity for scale-free \acrshort{QN}s under random node or edge breakdown in Fig.~\ref{fig:yookBreak_C}. As the upper bounds are not tight even in absence of the errors, we adopt a rescaling so that the general trend with $p$ can be compared. 

\subsection{Intentional attack}
\begin{figure}[t]
    \centering
    \includegraphics[width = 0.8\textwidth]{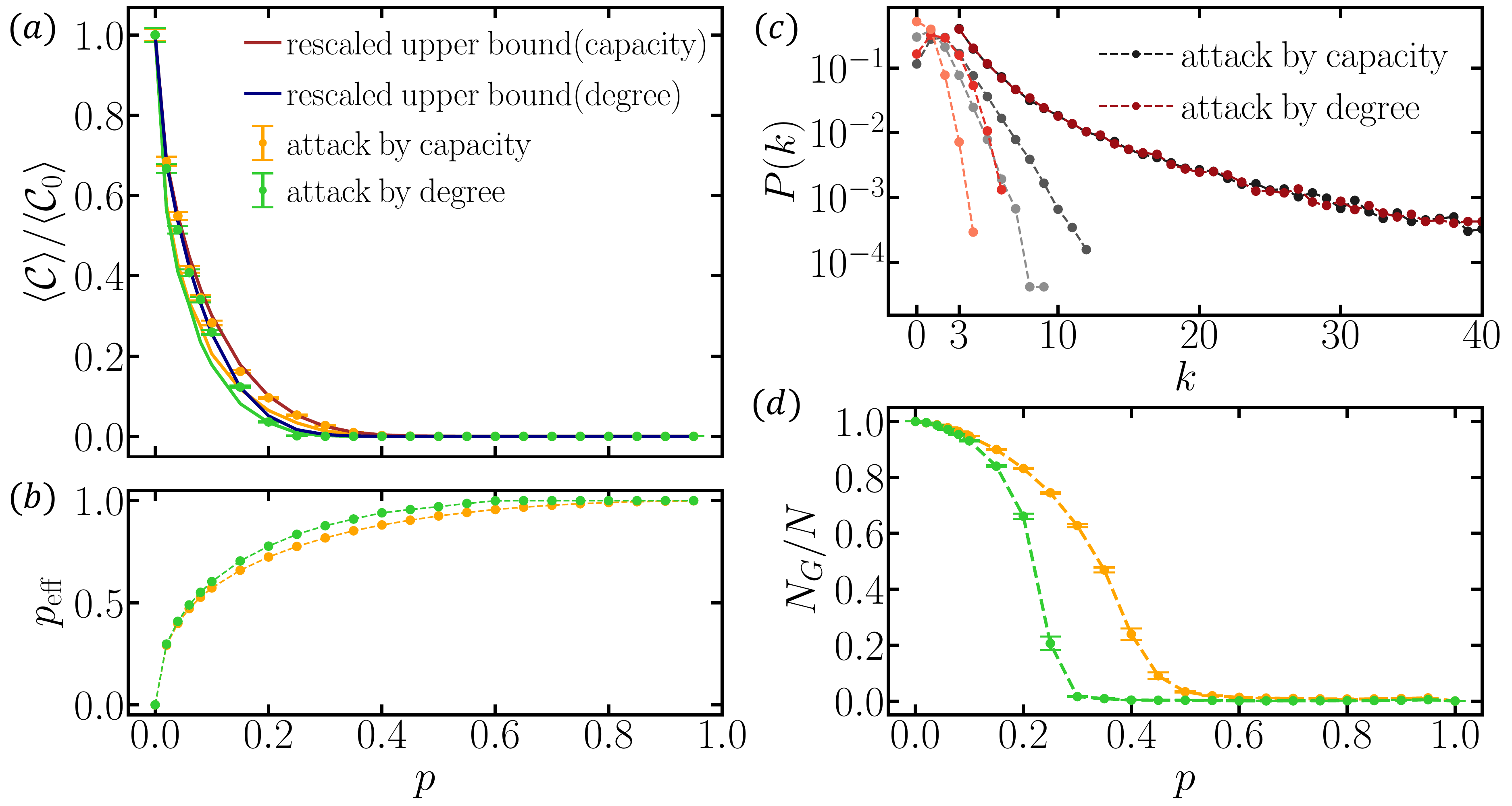}
    \caption{Scale-free \acrshort{QN} under attack by node capacity and node degree. (a) Normalized average end-to-end capacity $\braket{\mathcal{C}}/\braket{\mathcal{C}_0}$ vs fraction of attacked nodes $p$ with $R = 40$. Solid orange and green curves represent $\braket{\mathcal{C}}/\braket{\mathcal{C}_0}$ of scale-free \acrshort{QN} under random edge breakdown with corresponding probability $p_{\rm eff}$. Dark red and blue curves show the rescaled upper bound for edge breakdown Eq.~\eqref{eq:yook_nodebreak_C_bound} with probability $p_{\rm eff}$. \QZ{Dark red and blue curves show the rescaled upper bound for attack by capacity and degree separately according to Eq.~\ref{eq:yook_nodebreak_C_bound} with $p$ to be effective edge breakdown probability $p_{\rm eff}$ shown in (b).}
    (b) The ratio of edges that are removed through attack $p_{\rm eff}$ vs $p$. It shares the same legend as in (a).
    (c) Degree distribution $P(k)$ of scale-free \acrshort{QN} under attack with $p = 0, 0.2, 0.4$ (from dark to light for both black and red) at $R = 40$.
    (d) Relative size of giant component vs $p$ with $R = 40$. It shares same legend as (a).
    \label{fig:yookAttack_graph}
    }
\end{figure}

As for scale-free \acrshort{QN}s under attack, things are different. This can be immediately seen from the degree distribution $P(k)$ under attack, which gets closer to an exponential distribution rather than a power law as the fraction of removed nodes $p$ increases (see Fig.~\ref{fig:yookAttack_graph}(c)), indicating substantial structure changes. Indeed, the giant component disappears at a critical fraction much smaller than that under random breakdowns (see Fig.~\ref{fig:yookAttack_graph}(d)). The attack by node degree leads to a even sharper transition in both the degree distribution and the giant component transition than the attack by node capacity. 

Now we proceed to evaluate the average end-to-end capacity.
In contrast to the slow linear decrease in the Waxman case, the average capacity $\braket{\mathcal{C}}$ decreases with $p$ exponentially, shown in Fig.~\ref{fig:yookAttack_graph}(a) (details in Appendix). In particular, the capacity $\braket{\mathcal{C}}$ decays much faster than the giant component transition under both attacks, as we see $\braket{\mathcal{C}}/\braket{\mathcal{C}_0}$ is already below $0.1$ for $p\sim0.2$, which is within the parameter region supporting the existence of a giant component in the network. Besides, because the major contribution to the ensemble averaging of the capacity still comes from the giant component, Eq.~\ref{eq:yook_nodebreak_C_sol} still holds; The probability that a random edge is attached to an attacked node, $p_{\rm eff}$, is equal to the ratio of the number of edges removed during the attack $|\mathcal{E}_0|-|\mathcal{E}|$ to the original number of edges $|\mathcal{E}_0|$ prior to the attack~\cite{cohen2001breakdown}. Therefore, effects of the attack of fraction $p$ can be equivalently produced by the scale-free \acrshort{QN}s under random edge breakdown with an effective probability $p_{\rm eff}=1- {|\mathcal{E}|}/{|\mathcal{E}_0|}$ (Fig.~\ref{fig:yookAttack_graph}(b)). To verify the above intuition, we plot $\braket{\mathcal{C}}/\braket{\mathcal{C}_0}$ of scale-free \acrshort{QN}s under random edge breakdown with $p_{\rm eff}$ (orange and green curves) in Fig.~\ref{fig:yookAttack_graph}(a), which agrees well with results from numerical simulations of the attack (orange and green dots). By substituting all $p$ in Eq.~\ref{eq:yook_nodebreak_C_bound} (note that $N_G$ also depends on $p$) with $p_{\rm eff}$, we can obtain an upper bound for the scale-free \acrshort{QN}s under attack. In Fig.~\ref{fig:yookAttack_graph}(a), we see that despite the upper bounds being non-tight, the rescaled upper bounds (dark blue for attack by degree and dark red for attack by capacity) agree well with the numerical results (green and orange) in terms of the trend with the increasing $p$.

Despite the overall similarity between the two types of attacks under consideration, the attack by node degree leads to a slightly sharper transition in network properties and a lower capacity (see Appendix). This can be understood by the equivalence to random edge breakdowns with probability $p_{\rm eff}$. An attack on nodes is most effective when $p_{\rm eff}$ is maximum for a fraction of $p$ nodes being removed. Suppose we remove a set of nodes $A$, then we have
\QZ{
$ 
p_{\rm eff}\propto \sum_{\bm x\in A}D(\bm x),
$ 
}
which is maximized by the attack by node degree. Indeed, in Fig.~\ref{fig:yookAttack_graph}(b), we see the attack by degree gives a larger $p_{\rm eff}$ at any attack fraction. Therefore, assuming the mapping to edge breakdown, contrast to the intuition, attack by node degree is in fact the most effective attack for scale-free \acrshort{QN}s. 

\section*{Discussions}
In this paper, we examine the transition of end-to-end quantum capacity in Waxman and scale-free \acrshort{QN}s under errors, such as random breakdown and attack. For Waxman \acrshort{QN}s, we find a linear decrease of capacity caused by
the re-parameterization of Waxman model under errors. For scale-free \acrshort{QN}s, similar linear decrease happens for random breakdown; however, under attack by node capacity, the capacity of scale-free \acrshort{QN}s decay exponentially in the fraction of attack. All findings are supported by numerical evidence as well as analytical analyses. Our results give an insight of the capacity transition in \acrshort{QN}s with the growth of errors
and provide guidance on its design to cope with random breakdown and attack.

Parallel channels and redundancy design can generally increase the reliability of a network in practice. From our results, for the Waxman \acrshort{QN}s that are highly connected, parallel design would be helpful to protect the connectivity in quantum capacity in both random breakdown and attack. For scale-free \acrshort{QN}s against random breakdown, parallel and redundancy could still work, especially when the \acrshort{QN} is small-scale. However, the exponential decay under attack by node capacity indicates that such simple schemes are not efficient, and thus some extra methods are required to achieve a reliable and robust scale-free \acrshort{QN}.

After the completion of the manuscript, a related work appeared on arxiv~\cite{brito2}.

\begin{acknowledgments}
The project is supported by the Defense Advanced Research Projects Agency (DARPA) under Young Faculty Award (YFA) Grant No. N660012014029 and National Science Foundation Engineering Research Center for Quantum Networks Grant No. 1941583. Q.Z. acknowledges Craig M. Berge Dean's Faculty Fellowship of University of Arizona. 
\end{acknowledgments}

\appendix

\section{Degree distribution and giant component transition}
\label{app:degree}
In this section, we provide more details about the degree distribution and giant component transition of both the Waxman and scale-free networks.

\begin{figure}[t]
    \centering
    \includegraphics[width=1\textwidth]{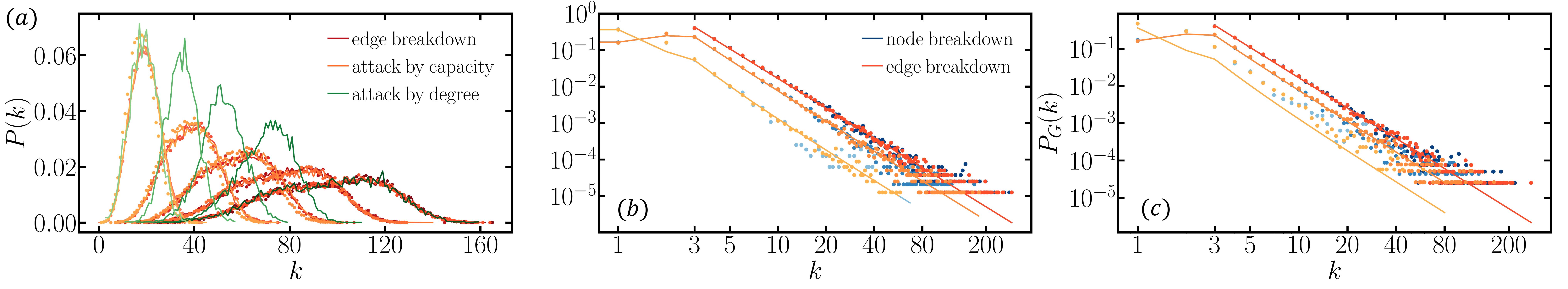}
    \caption{(a) \QZ{Degree distribution $P(k)$ of Waxman model under random edge breakdown (red), attack by capacity (yellow) and attack by degree (green) with $\alpha = 0.1$.} Dots and curves with black and blue represent $P(k)$ of Waxman model under real errors and the equivalent model. Colors from dark to light correspond $p = 0, 0.2, 0.4, 0.6, 0.8$. (b), (c) Degree distribution of the scale-free network (b) and the giant component of it (c) under random node (black) and edge (blue) breakdown with $R = 40$. Solid curves in (b) represent the analytical expression of degree distribution \cite{cohen2000resilience, martin2006random}, with power-law exponent $\nu\sim 2.7$. Solid curves in (c) is the same as the ones in (b) as an approximation. Colors from dark to light correspond to $p = 0, 0.4, 0.8$.
    }
    \label{fig:degree_supp}
\end{figure}

When a Waxman model undergoes random edge breakdown, its edge density decreases while the other parameters are unchanged. By comparing the degree distribution, we show that it is statistically equivalent to the Waxman model with edge connection probability in Eq.~\eqref{eq:edge_probability}. As we see in Fig.~\ref{fig:degree_supp}(a), the degree distribution of Waxman model under fraction-$p$ attack by node capacity agrees well with a Waxman model with no error but $N = N_0(1-p)$ nodes; Moreover, as we already see in Fig.~\ref{fig:degree}(a), the degree distribution of a Waxman model under random node breakdown is also identical to that of a Waxman model with no error but $N = N_0(1-p)$ nodes; Therefore, a Waxman model, regardless of under probability-$p$ random breakdown or fraction-$p$ attack by capacity, is equivalent to a Waxman model with no error but a fewer $N = N_0(1-p)$ number of nodes. However, under attack by node degree, the degree distribution of a Waxman model does not agree with the other cases, as shown by the red lines in Fig.~\ref{fig:degree_supp}(a).

Without any breakdown, the degree distribution of scale-free networks shows a power-law, $P(k)\sim k^{-\nu}$. As shown in Fig.~\ref{fig:degree_supp}(b), the power-law persists under random breakdown, although a large portion of nodes has a degree $k<m_0$ due to breakdown~\cite{cohen2000resilience, martin2006random}. To better understand the structure, we also focus on the giant component and study its degree distribution in Fig.~\ref{fig:degree_supp}(c), where we see similar power-law persisting. Qualitative agreement still holds despite deviations happen for larger $p$.

\begin{figure}[t]
    \centering
    \includegraphics[width = 0.7\textwidth]{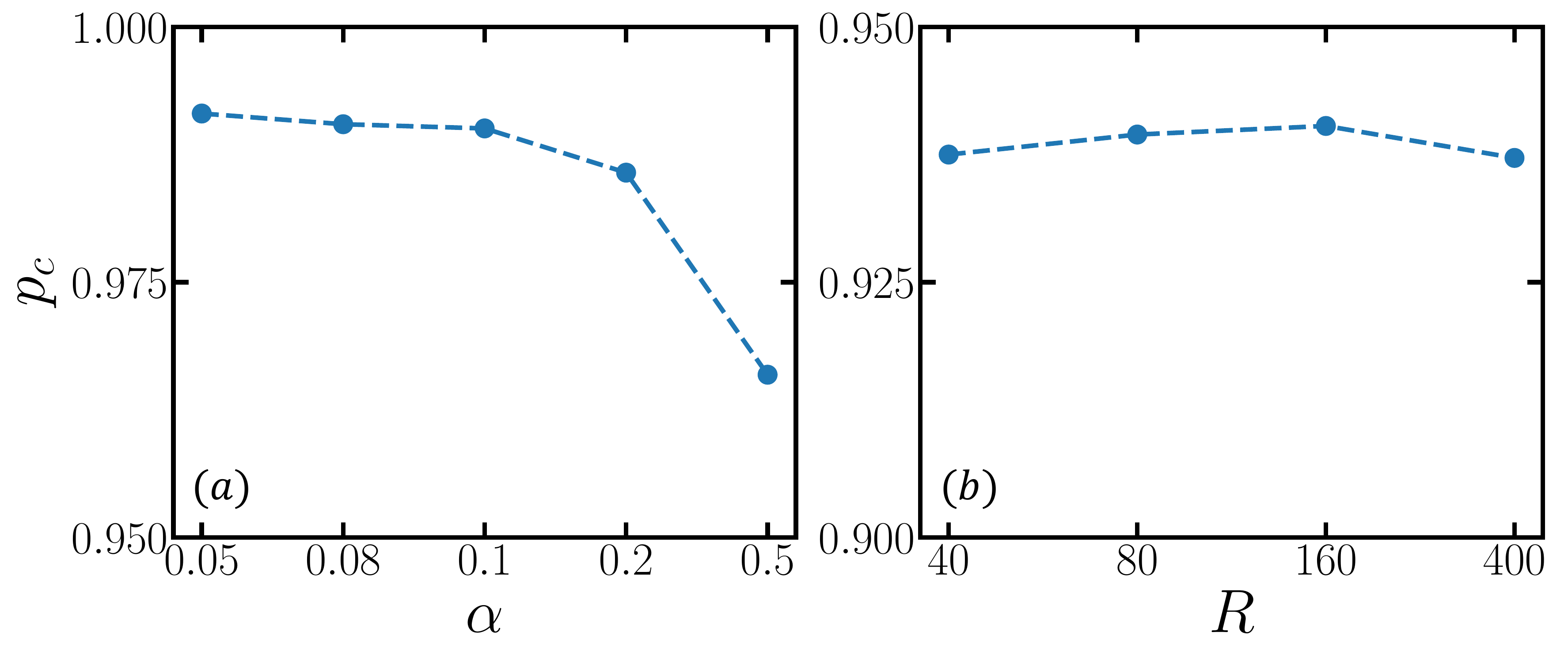}
    \caption{\QZ{Critical probability $p_c$ for Waxman networks (a) with different $\alpha$ and scale-free networks (b) with different $R$.}}
    \label{fig:kappa}
\end{figure}

The critical probability $p_c$ of giant component transition for a Waxman model or a scale-free network under random breakdown can be directly solved via~\cite{cohen2000resilience}
\begin{equation}
    p_c = 1-\frac{1}{\braket{k_0^2}/\braket{k_0} - 1}
\end{equation}
where $\braket{k_0^2}$ and $\braket{k_0}$ are the average of degree squared and the average degree of unbroken network. From the degree distributions we can obtain the values of $\braket{k_0^2}/\braket{k_0}$ in Fig.~\ref{fig:kappa}; as the values are large, the critical probability is close to unity.

\section{Average end-to-end capacity upper bounds}
\subsection{Waxman \acrshort{QN}s}
From Ref.~\cite{ourpaper}, we have the upper bound of average end-to-end capacity of a Waxman \acrshort{QN} with connection probability $\Pi_{\rm W}$ as
\begin{eqnarray}
&\braket{\mathcal{C}_N(\bm x)} = \rho \zeta_{\rm W},
\label{UB_W1}
\\
&\zeta_W =\frac{1}{|\Omega_R|}\int_{\Omega_R}d^2\bm{x}\int_{\Omega_R}d^2\bm{x}^\prime \Pi_{\rm W}(\bm x, \bm x^\prime)\mathcal{C}_E(E_{\bm x,\bm x^\prime})
,
\label{UB_W2}
\end{eqnarray}
where $\Omega_R$ denotes the $2R\times 2R$ region, $|\Omega_R|=4R^2$ and $\rho$ is the density of nodes.

Under random node breakdown of probability $p$ or attack by capacity of fraction $p$, we can simply replace $\rho=\rho_0 (1-p)$ in Eq.~\eqref{UB_W1} to obtain the upper bound $\braket{\mathcal{C}_N(\bm x)}=(1-p)\rho_0 \zeta_W$.

When the Waxman \acrshort{QN} undergoes random edge breakdown, we simply replace the connection probability $\Pi_W$ in Eq.~\eqref{UB_W2} with $\Pi_{\rm W}^{\rm E}$ in Eq.~\eqref{eq:edge_probability}, leading to the same upper bound $\braket{\mathcal{C}_N(\bm x)}=(1-p)\rho_0 \zeta_W$.

\subsection{Scale-free \acrshort{QN}s} 
From Ref.~\cite{ourpaper}, we have the upper bound for the scale-free \acrshort{QN}s as
\begin{eqnarray}
&\braket{\mathcal{C}_N\left(\bm x\right)}=2m\zeta_{\rm SF},
\\
&\zeta_{\rm SF}=
\frac{1}{ A^\prime}\int_{\Omega_R} d^2\bm  x \int_{\Omega_R} d^2 \bm x^\prime    \braket{\frac{1}{\QZ{d\left(\bm x, \bm x^\prime\right)}} \mathcal{C}_E(E_{\bm x,\bm x^\prime})},
\\
&A^\prime=\int_{\Omega_R} d^2\bm  x \int_{\Omega_R} d^2 \bm x^\prime    \braket{\frac{1}{\QZ{d\left(\bm x, \bm x^\prime\right)}}}.
\end{eqnarray}

As only the average degree comes into play, when the node or edge breakdown leads to a new average degree $m=2m_0(1-p)$, we have the upper bound $\braket{\mathcal{C}\left(\bm x\right)}= 2m_0(1-p)\zeta_{\rm SF}$.

\section{More details for the average capacity}

\subsection{Waxman \acrshort{QN}s}
\begin{figure}[t]
    \centering
    \includegraphics[width = 0.6\textwidth]{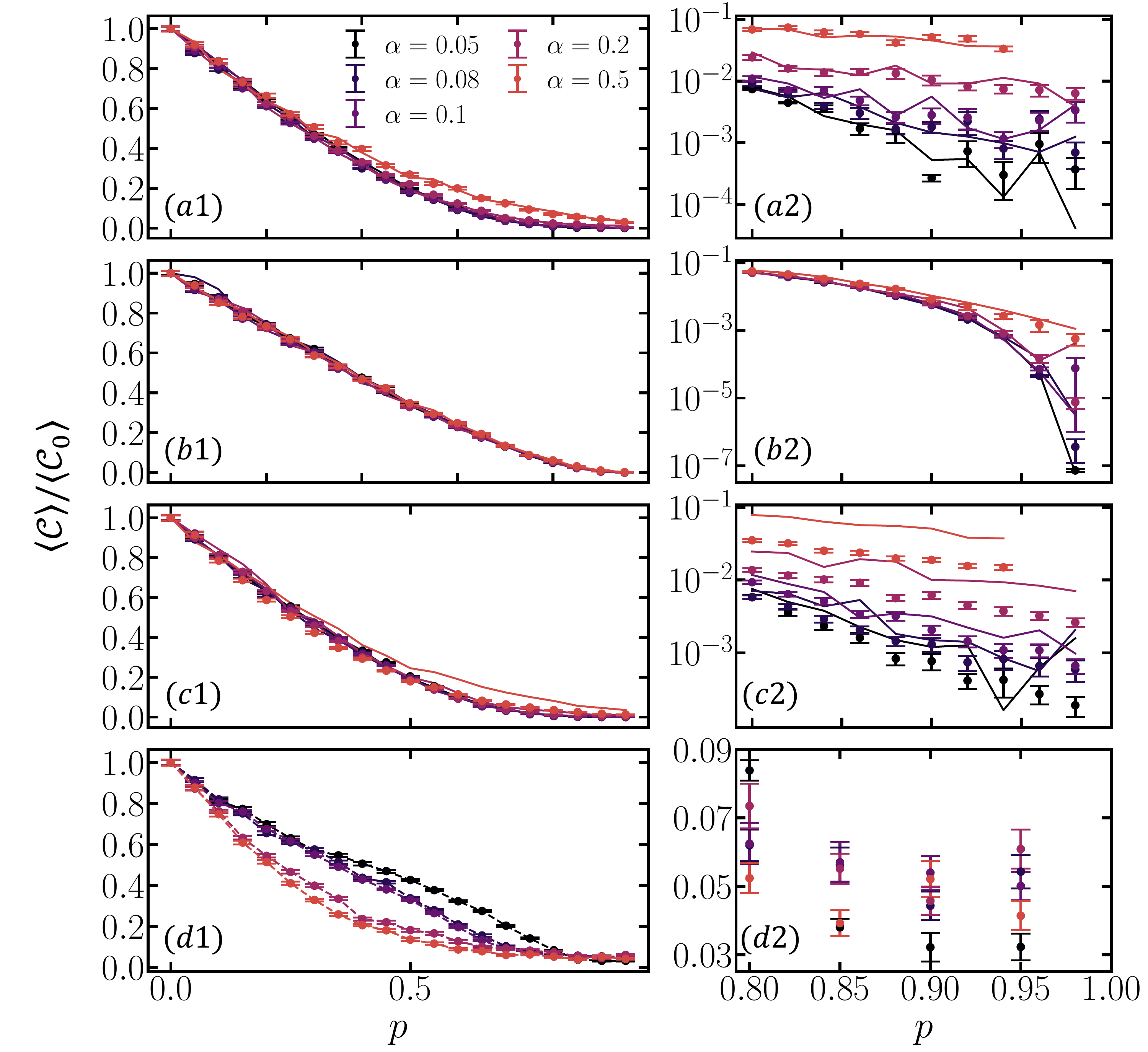}
    \caption{Normalized average end-to-end capacity $\braket{\mathcal{C}}/\braket{\mathcal{C}_0}$ vs $p$ with different $\alpha$. \QZ{From (a) to (d) we plot the capacity of Waxman \acrshort{QN} under random node breakdown, random edge breakdown, attack by node capacity and attack by node degree}. Subfigures in the right column show a zoom in of average capacity when $p\geq 0.8$. \QZ{Dots with error bars represent capacity under random breakdown or attack and solid curves in (a)-(c) show capacity from Waxman model with a re-parameterization.}}
    \label{fig:waxman_C_supp}
\end{figure}

In Fig.~\ref{fig:waxman_C_supp}, we show that the average capacity of Waxman \acrshort{QN}s under random breakdown and attack by capacity agrees among different $\alpha$, and the re-parameterization of Waxman model can also provide a good description of capacity even when $p$ is large. \QZ{Note that for $\alpha=0.5$, average capacity decays slightly faster under attack by capacity than random breakdowns. This is because the system size and initial number of nodes are both small, the node capacity is not as uniform as other cases which stronger effect in attack than breakdown.}

\begin{figure}
    \centering
    \includegraphics[width = 0.4\textwidth]{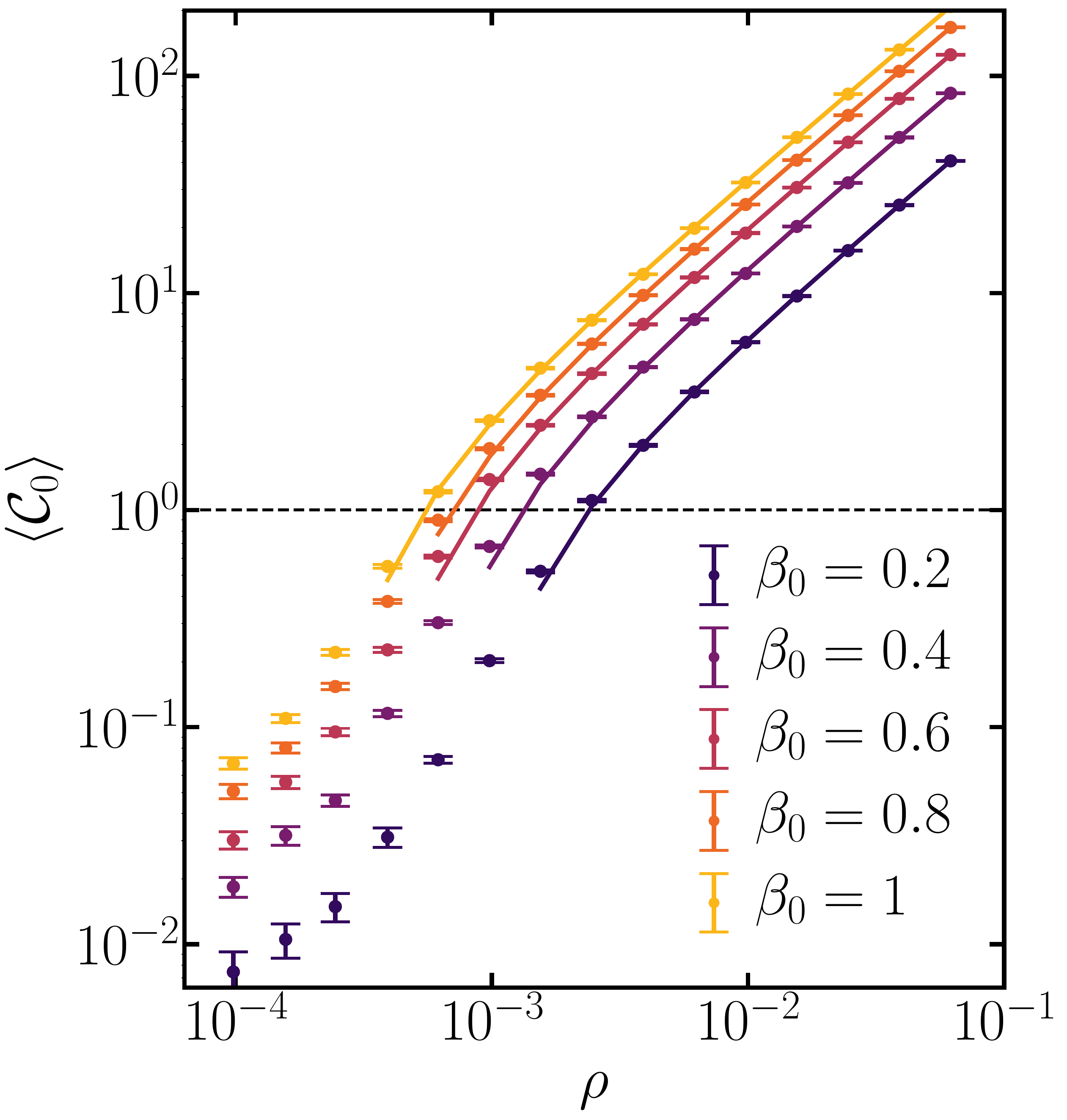}
    \caption{Average end-to-end capacity $\braket{\mathcal{C}_0}$ vs node density $\rho$ with different $\beta_0$ and $\alpha = 0.5$. Solid curves give a linear fitting for $\braket{\mathcal{C}_0}$ vs $\rho$.}
    \label{fig:waxman_C_beta}
\end{figure}

We also plot how the capacity $\braket{\mathcal{C}_0}$ behaves with different choices of $\beta$ vs density of nodes $\rho$ in Waxman \acrshort{QN}s \QZ{(see Fig.~\ref{fig:waxman_C_beta})}. The gradient $d\braket{\mathcal{C}_0}/d\rho$ increases with $\beta_0$, which indicates the comparably slow decay of capacity under edge breakdown.

\begin{figure}[t]
    \centering
    \includegraphics[width = 0.6\textwidth]{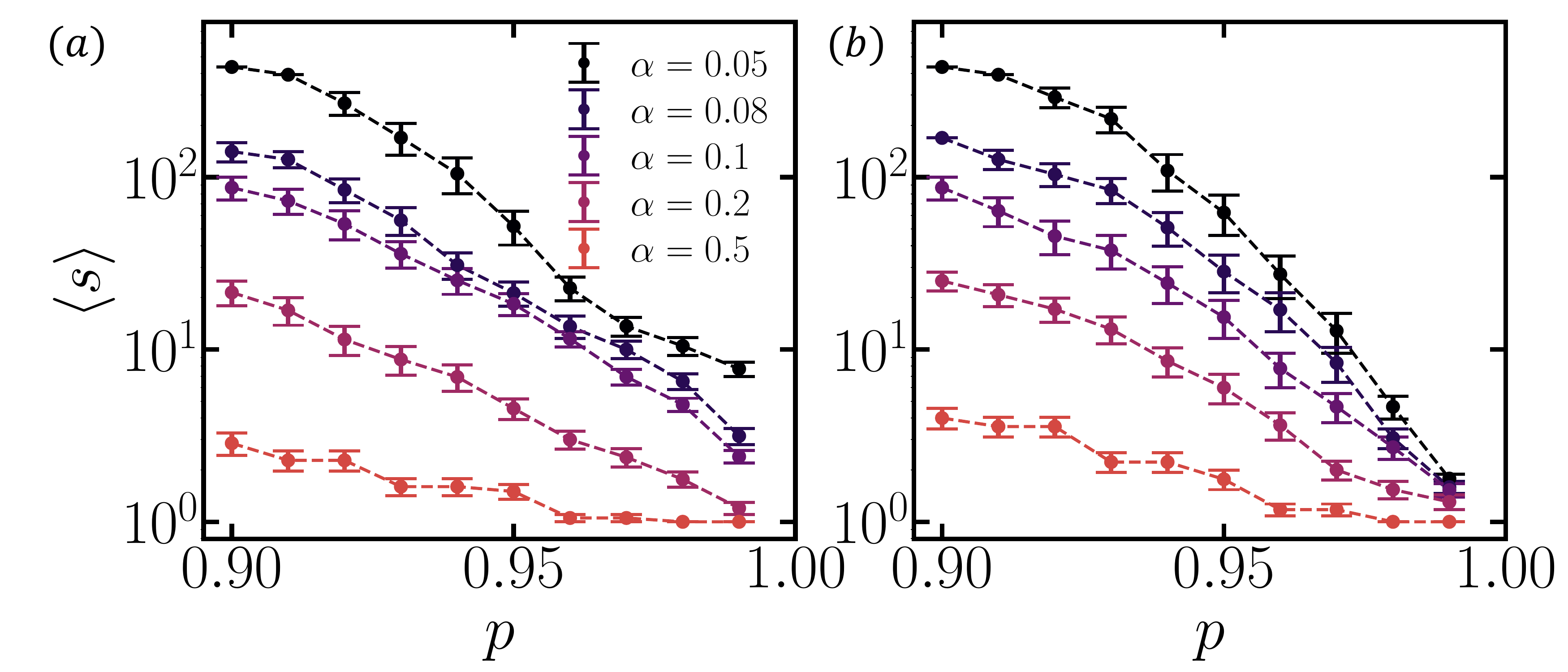}
    \caption{Average size of the connected component $\braket{s}$ for Waxman \acrshort{QN}s under attack by node degree (b) and attack by node capacity (c) when $p$ is large. They shares the same legend.}
    \label{fig:waxman_attackD_C}
\end{figure}

For the Waxman \acrshort{QN}s under attack by node degree \QZ{(plotted in Fig.~\ref{fig:waxman_C_supp}(d1))}, although $\braket{\mathcal{C}}/\braket{\mathcal{C}_0}$ varies with different $\alpha$, generally the capacity still decays linearly and slightly deviates from the results of random breakdown and attack by capacity (see Fig.~\ref{fig:waxman_C}). When $p$ is large, the average size of connected components $\braket{s}$ is larger under attack by degree than attack by capacity, as shown in \QZ{Fig.~\ref{fig:waxman_attackD_C}(a), (b)}. Those connected clusters preserve a non-negligible amount of capacity, leading to the nonzero tail in \QZ{Fig.~\ref{fig:waxman_C_supp}(d1), which is also seen in Fig.\ref{fig:waxman_C} for the attack by node degree case. We want to emphasize that here we consider the large number of nodes case, where the total number of the leftover nodes $N_0(1-p)$ is still large.}

\subsection{Scale-free \acrshort{QN}s}

\begin{figure}[t]
    \centering
    \includegraphics[width = 0.8\textwidth]{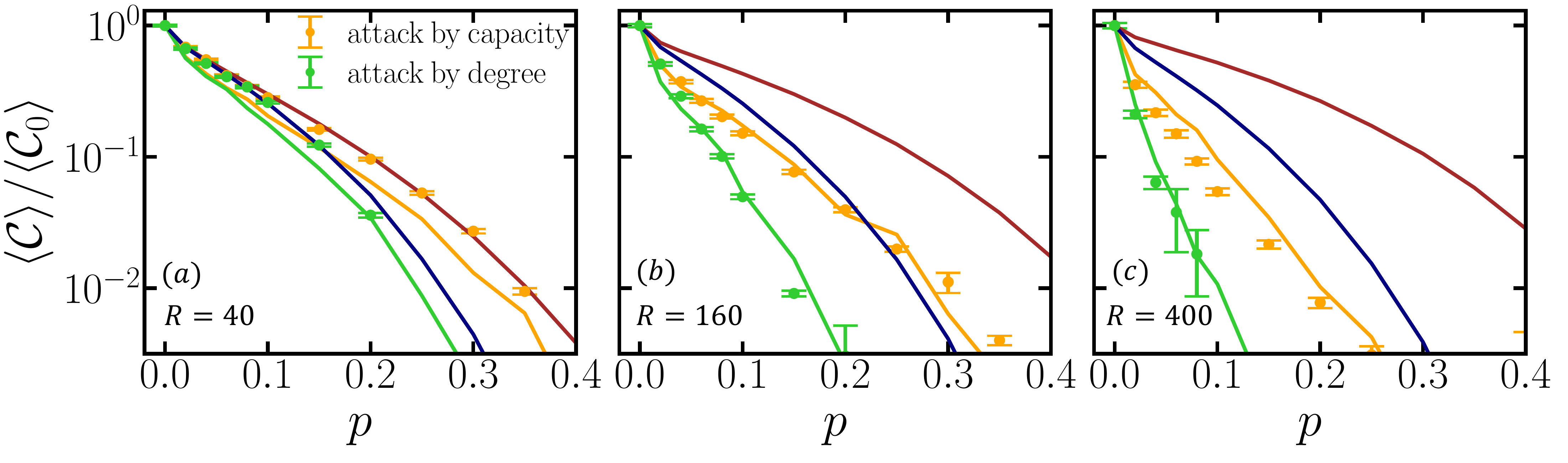}
    \caption{$\braket{\mathcal{C}}/\braket{\mathcal{C}_0}$ vs $p$ of scale-free \acrshort{QN}s under attack by node capacity (orange) and node degree (green) with $R = 40, 160, 400$ in (a), (b), (c) correspondingly. Solid orange and green curves represent the random edge breakdown results with corresponding probability $p_{\rm eff}$. Dark red and blue curves are the rescaled upper bound.
    }
    \label{fig:yookAttack_C_supp}
\end{figure}

As the scale-free network under random breakdown, from the definition of average end-to-end capacity, we have
\begin{eqnarray}
    \braket{\mathcal{C}} &= \frac{1}{\binom{N}{2}}\sum_{\bm{x}, \bm{x}^\prime}\mathcal{C}(\bm{x},\bm{x}^\prime)\\
    &= \frac{1}{\binom{N}{2}}\left(\sum_{\bm{x}, \bm{x}^\prime \in G}\mathcal{C}(\bm{x},\bm{x}^\prime) + \sum_{\rm else}\mathcal{C}(\bm{x},\bm{x}^\prime) \right) 
    \label{C2}
    \\
    &= \frac{1}{\binom{N}{2}}\sum_{\bm{x}, \bm{x}^\prime \in G}\mathcal{C}(\bm{x},\bm{x}^\prime)\left(1+O\left(\frac{1}{N_G^2}\right)\right)
    \label{C3}
    \\
    &= \frac{\binom{N_G}{2}\braket{\mathcal{C}_G}}{\binom{N}{2}},
\end{eqnarray}
where $\braket{\mathcal{C}_G}$ is the average end-to-end capacity of the giant component in the network, and number of nodes $N = N_0(1-p)$ for node breakdown while $N = N_0$ for edge breakdown. \QZ{The two summations in Eq.~\eqref{C2} are the end-to-end capacity contributed from different components, the first item from the large component and the second item from all other small components. Following the analyses in Sec.~\ref{sec:rb}, we can approximate contributions from those small components by $\sum_{\rm else}\mathcal{C}(\bm{x}, \bm{x}^\prime)\sim  \braket{\mathcal{C}_G}\sum_k \binom{N_k}{2}$, where $N_k$ is the number of nodes in the $k$th component and we sum over all components $k$ excluding the largest component.
We can then obtain $\braket{\mathcal{C}_G}\sum_k \binom{N_k}{2} =  \sum_{\bm{x},\bm{x}^\prime \in G} \mathcal{C}(\bm{x},\bm{x}^\prime) \sum_k \binom{N_k}{2}/\binom{N_G}{2} \sim \sum_{\bm{x},\bm{x}^\prime \in G} \mathcal{C}(\bm{x},\bm{x}^\prime) O(1/N_G^2)$, which is Eq.(C.3).}

For scale-free \acrshort{QN}s under attack, we can directly see the exponential decay of $\braket{\mathcal{C}}/\braket{\mathcal{C}_0}$ from Fig.~\ref{fig:yookAttack_C_supp}, which can be well characterized by the equivalent random edge breakdown with probability $p_{\rm eff}$. The rescaled upper bound (dark blue and red curves) also show the exponential decay of capacity. Compared to node capacity attack, the node degree attack is more effective for scale-free \acrshort{QN} in terms of the reduction on average capacity.

\section{More details on numerical simulation}

\QZ{
We utilize python-igraph~\cite{igraph} for numerical simulation of \acrshort{QN}s. To evaluate the ensemble-averaged properties of \acrshort{QN}s, we simulate $10$ random \acrshort{QN}s for each case, and randomly sample $200$ pairs of nodes for calculation to represent average end-to-end capacity for each graph. In Fig.~\ref{fig:waxman_C}, we do an numerical experiment on 10 graphs, each with $500$ pairs to make it smooth on $N$ re-parameterization. When nodes are removed, the number of pairs to be sampled in each graph is also reduced.
}

\end{document}